\documentclass[preprint,showpacs,showkeys,preprintnumbers,amsmath,amssymb]{revtex4}

\usepackage{graphicx}

\usepackage{dcolumn}
\usepackage{bm}

\begin{document}



\title{Observation of Anti-correlation between Scintillation and Ionization for MeV Gamma-Rays in Liquid Xenon}


\author{E. Aprile}
\author{K.L. Giboni}
\author{P. Majewski\footnote{Current address: Department of Physics and Astronomy, University of Sheffield, UK.}}
\author{K. Ni\footnote{Current address: Physics Department, Yale University, New Haven, CT 06511.}}
\author{M. Yamashita}
\affiliation{Physics Department and Columbia Astrophysics Laboratory, Columbia University, New York, New York 10027}

\date{\today}

\begin{abstract}
A strong anti-correlation between  ionization and scintillation signals produced by MeV $\gamma$-rays in liquid xenon has been measured and used to improve the energy resolution by combining the two signals.  The improvement is explained by reduced electron-ion recombination fluctuations of the combined signal compared to fluctuations of the individual signals.  Simultaneous measurements of ionization and scintillation signals were carried out with $^{137}$Cs, $^{22}$Na and $^{60}$Co $\gamma$ rays, as a function of electric field in the liquid. A resolution of 1.7\%($\sigma$) at 662 keV was measured at 1 kV/cm, significantly better than the resolution from either scintillation or ionization alone. A detailed analysis indicates that further improvement to less than 1\%($\sigma$) is possible with higher light collection efficiency and lower electronic noise.

\end{abstract}

\maketitle

\section{Introduction}
\label{}
Liquid xenon (LXe) is an excellent medium for  radiation detection, with high stopping power, good ionization and scintillation yields. Currently liquid xenon detectors are being developed for several fundamental particle physics experiments, from neutrinoless double beta decay \cite{EXO} and dark matter weakly interactive massive particles (WIMPs) detection \cite{XENON, Alner:2007ja, Akimov:2006qw, Yamashita:Xenon01}, to spectroscopy and imaging of gamma-rays in physics, astrophysics and nuclear medicine \cite{MEG, Aprile:01, Chepel:97PET, Doke:06PET}. A more precise energy measurement than currently demonstrated with liquid xenon ionization and scintillation detectors would largely benefit all these experiments. The best experimental energy resolution is not only orders of magnitude worse than that expected from the Fano factor \cite{Fano} but  even worse than that predicted by the Poisson statistics, based on the measured W-value of 15.6 eV \cite{Takahashi:75}, as average energy to produce an electron-in pair. The reason for the discrepancy is yet to be fully understood but fluctuations in electron-ion pair recombination rate are known to play a dominant role. 
Both electron-ion pairs and excitons are produced by the passage of an ionizing particle in liquid xenon. In the presence of an electric field, some of the electron-ion pairs are separated before recombination, providing the charge signal as electrons drift freely in the field (positive ions do not contribute as their drift velocity is several orders of magnitude slower). Recombination of the remaining electron-ion pairs lead to excited xenon molecules, $\rm{Xe_{2}^{\ast}}$. Excitons that are directly produced by the incident particle also become $\rm{Xe_{2}^{\ast}}$ molecules. De-excitation of these molecules to the ground state, $\rm{Xe_{2}^{\ast}} \rightarrow \rm{2Xe} + \mathit{h\nu}$, produce scintillation photons with a wavelength of 178 nm~\cite{Jortner:65}. The ionization and scintillation signals in liquid xenon are thus complementary and anti-correlated as the suppression of recombination by the external field  results in more free electrons and less scintillation photons. 
This anti-correlation was first observed by Kubota {\it et al.}~\cite{Kubota:78}. Large fluctuations in the number of collected electrons due to their reduction by recombination lead to poor energy resolution of the ionization signal. A way to increase the ionization signal and thus the energy resolution via the photo-ionization effect in LXe doped with triemethylamine (TEA) yielded good results, but only at low electric fields \cite{Ichinose:92}. Another way to improve the energy resolution is to reduce recombination fluctuations by combining ionization and scintillation signals. Since recombination also produce scintillation photons, fluctuations of the combined signal should be reduced.  This was originally suggested by  Ypsilantis {\it et al.}  many years ago \cite{Seguinot:92}, but the simultaneous detection of scintillation and ionization in LXe has been hard to realize because of the difficulty to efficiently detect VUV light under the constraints of efficient charge collection. 
In the Liquid Xenon Gamma-Ray Imaging Telescope (LXeGRIT) \cite{Aprile:01}, which we developed for Compton imaging of cosmic $\gamma$-rays, both ionization and scintillation were detected, but the fast scintillation signal merely provided the event trigger, while the ionization signal provided the energy measurement, with a resolution of 4.2\%($\sigma$) at 1 MeV. The fair resolution has been a major limitation of the LXe time projection chamber (TPC)  technology for astrophysics. In recent years, the development of VUV-sensitive photomultiplier tubes (PMTs), capable to operate directly in the cryogenic liquid and to withstand overpressure up to 5 atm, has enabled to significantly improve the Xe scintillation light collection efficiency with good uniformity across the liquid volume. A light readout based on these novel PMTs, coupled with a lower-noise charge readout, was proposed to measure both signals event-by-event and thus to improve the energy resolution and the Compton imaging of the LXeGRIT telescope \cite{Aprile:03}. The work presented here was carried out with two of the first PMTs developed for operation in LXe and is our first attempt in this direction. Further optimization of these VUV PMTs has continued, driven largely by our XENON dark matter detector development~\cite{XENON, Aprile:05a}. These improved PMTs, along with other VUV sensor technologies such as Large Area Avalanche Photodiodes (LAAPDs) and Si photomultipliers (SiPMs), which we are also testing for LXe scintillation detection~\cite{APD, SiPM}, promise further energy resolution improvement.

\section{Experimental Set-up and Signals}

\begin{figure}
\centering
\includegraphics[width=0.4\textwidth]{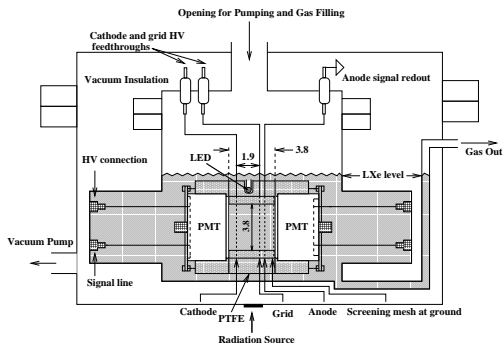}
\caption{The detector schematics (see text for details).}
\label{schematics}
\end{figure}

\begin{figure}
\centering
\includegraphics[width=0.45\textwidth]{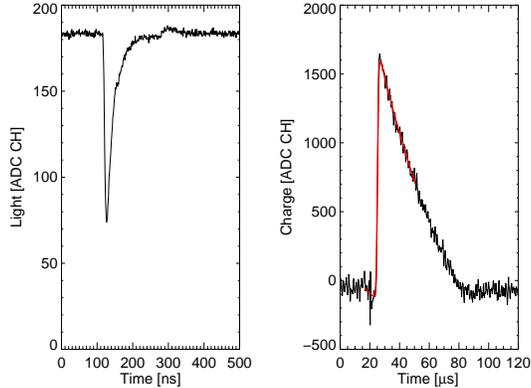}
\caption{Waveforms of scintillation signal (left, sum of two PMTs) and ionization signal (right) of a 662 keV \(\gamma\)-ray event from $^{137}$Cs at 1 kV/cm drift field.  }
\label{signal_wave}
\end{figure}

The detector used for this study is a gridded ionization chamber with two VUV sensitive PMTs (two-inch diameter Hamamatsu R9288) viewing the sensitive liquid xenon volume from the anode and cathode side. The two PMTs, and the transparent meshes serving as anode, cathode and shielding grid, are mounted  in a structure made of Teflon for its VUV reflectivity~\cite{Yamashita:04} (Fig. \ref{schematics}). The drift gap, between cathode and grid, is 1.9 cm  while the distance between grid and anode is 3 mm. Separate high voltage is supplied to the cathode and the grid, keeping a ratio between the field in the drift gap and the field in the collection gap such as to  maximize electron transmission through the grid. The  electrons collected on the anode are detected by a charge sensitive amplifier (ClearPulse Model 580). The charge signal is subsequently  digitized with 10 bit resolution and a sampling time of 200~ns (LeCroy Model 2262). The scintillation signal from each of the two PMTs is recorded with a digital oscilloscope (LeCroy Model LT374) with 1 ns sampling time. The time difference between the scintillation and ionization signals is the electron drift time which gives the event depth-of-interaction information. The coincidence of the two PMT signals is used as event trigger. Fig. \ref{signal_wave} shows the scintillation and ionization waveforms recorded at 1 kV/cm for a 662 keV \(\gamma\)-ray event from \(^{137}\)Cs. The number of photoelectrons, $N_{pe}$, detected by the PMTs is calculated based on the gain calibration with a light emitting diode (LED). The charge waveform is well described by the Fermi-Dirac threshold function in equation \ref{eq:fd_func}, as shown in \cite{Aprile:02}The pulse height $A$, drift time $t_d$, rise time $t_r$ and fall time $t_f$ are determined from fitting equation \ref{eq:fd_func} to the charge waveform. A known test pulse is used to calibrate the charge readout system,  and the number of collected electrons, $N_e$, is calculated from the pulse height of the charge waveform. 

\begin{equation}
\label{eq:fd_func}
Q(t) = A \cdot \frac{e^{-(t-t_d)/t_r}}{1+e^{-(t-t_d)/t_f}}
\end{equation}

The Teflon structure holding the PMTs and the meshes is mounted in a stainless steel vessel filled with liquid xenon at about --95$\rm ^o$C during the experiment. A vacuum cryostat surrounds the vessel for thermal insulation.  The xenon gas filling and purification system, as well as the ``cold finger" system used for this set-up is described in a previous publication\cite{Aprile:02}. The set-up was modified for these measurements by  adding a gas recirculation system \cite{Aprile:05a} in order to purify the xenon continuously until sufficient charge collection is reached.

\section{Results and Discussion}

\subsection{Field dependence of scintillation and ionization}
\begin{figure}
\centering
\includegraphics[width=0.45\textwidth]{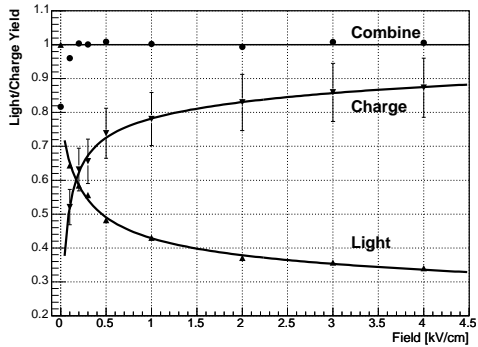}
\caption{Light and charge yield as a function of drift field for 662 keV $\gamma$-rays from $^{137}$Cs. The uncertainty for charge measurement is due to the uncertainty in pre-amplifier calibration. The light yield is relative to that at zero field, for which the systematic uncertainty is negligible.}
\label{ly_field}
\end{figure}

Fig. \ref{ly_field} shows our measurement of the field dependence up to 4 kV/cm of the light and charge yield for 662 keV $\gamma$ rays from $^{137}$Cs. With increasing drift field, the charge yield increases, while the light yield decreases. This behavior has been known for a long time, and was originally reported in \cite{Kubota:78}. 

A parametrization of the field dependence of the light yield, $S(E)/S_0$, was proposed by Doke {\it et al}. \cite{Doke:02} introducing the model of escaping electrons to explain the scintillation light reduction at low LET(linear energy transfer). In this parametrization, expressed by equation~\ref{eq:svs0}, the light yield $S(E)$ at drift field $E$, normalized by the light yield at zero field, $S_0$, depends on the charge yield $Q(E)$ normalized to the charge at infinite field, $Q_0$, and on the ratio $N_{ex}/N_i$ of the number of excitons and ion pairs produced by a $\gamma$-ray.  
$\chi$ is the fraction of escaping electrons, i.e. the fraction of $N_i$ electrons which do not recombine with positive ions for an extended time ($>$ms) even at zero field, when the probability of recombination is highest.
 \begin{equation}
\label{eq:svs0}
\frac{S(E)}{S_0} = \frac{1+N_{ex}/N_i-Q(E)/Q_0}{1+N_{ex}/N_i-\chi}
\end{equation}
 $N_{ex}/N_i$ and $\chi$ can be determined from a fit of equation \ref{eq:svs0} to the charge and light yields data, knowing $Q_0$ which is given by $E_\gamma/W$, where $E_\gamma$ is the $\gamma$-ray deposited energy and $W = 15.6$ eV \cite{Takahashi:75} is the average energy required to produce an electron-ion pair in liquid xenon.  
 
\begin{figure}
\centering
\includegraphics[width=0.45\textwidth]{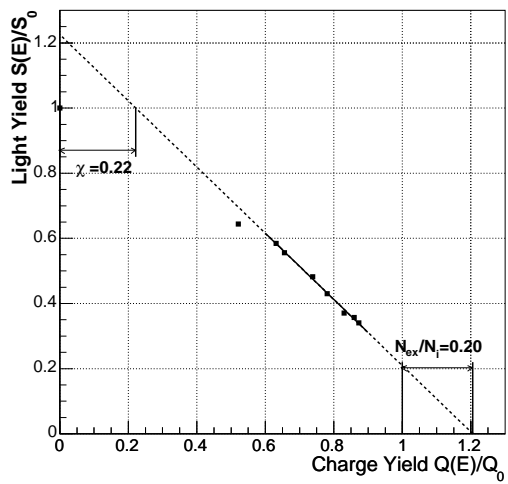}
\caption{Correlation between light yield and charge yield for 662 keV $\gamma$ rays.}
\label{ly_correlation}
\end{figure}

Fig. \ref{ly_correlation} shows the result of such a fit to our 662 keV data which gives $N_{ex}/N_i = 0.20\pm0.13$ and $\chi = 0.22\pm0.02$. The errors are from the uncertainty on charge collection only. A ratio of 0.06 for $N_{ex}/N_i$ was originally estimated from the optical approximation, using the absorption spectrum of solid rare gases \cite{Miyajima:74}. In  \cite{Doke:02}, $N_{ex}/N_i = 0.20$ and $\chi = 0.43$ as estimated from 1 MeV conversion electrons data in LXe. This $N_{ex}/N_i$ value is consistent with  that obtained from our data. The difference in  $\chi$ might be due to the limited range of electric fields used in our study.  


The charge and light signals can be combined by the following equation,
\begin{equation}
\label{eq:cvc0}
C(E) = a\frac{Q(E)}{Q_0}+b\frac{S(E)}{S_0}
\end{equation}
with $a = 1/(1+N_{ex}/N_i)$ and $b = 1-a\chi$, which gives a constant $C(E) = 1$, regardless of applied field. The proportion of charge and light is different at different fields but their sum is constant, as verified by our data in  in Fig. \ref{ly_field}. Note that at very low fields, equation \ref{eq:svs0}  and \ref{eq:cvc0} are not valid as escaping electrons are not fully collected.

\subsection{Combined Energy from Scintillation and Ionization}
The observed field dependent anti-correlation between  charge and light signals and its linear relationship offers  a way to improve the energy resolution by combining the two signals with proper coefficients. This was first shown in a measurement of energy loss of relativistic La ions in liquid argon \cite{Crawford:87}. More recently, Conti et al. \cite{Conti:03} applied the same method to improve the energy measurement of relativistic electrons in a liquid xenon detector using a single UV PMT to detect the scintillation signal.  For 570 keV $\gamma$-rays from $^{207}$Bi, an energy resolution of 3\%($\sigma$) was measured at 4 kV/cm, by combining the charge and light signals. In our study, the improved light collection efficiency with two PMTs immersed in the liquid gives even better results.

\begin{figure}
\centering
\includegraphics[width=0.5\textwidth]{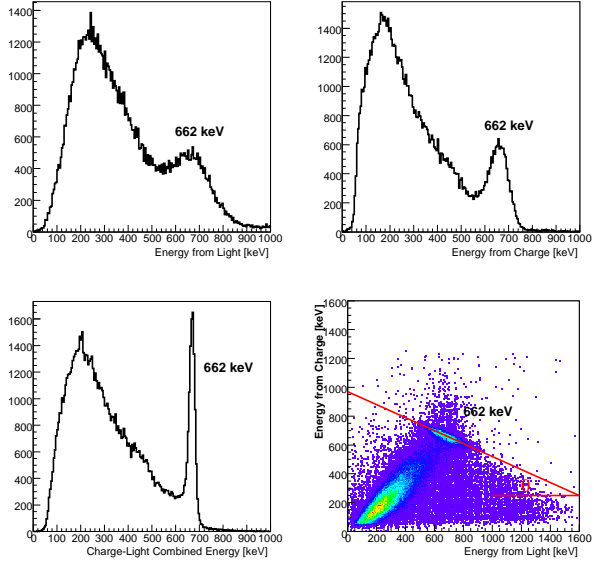}
\caption{Energy spectra of $^{137}$Cs 662 keV $\gamma$-rays at 1 kV/cm drift field in liquid xenon. The top two plots are from scintillation and ionization separately. The strong charge-light anti-correlation is shown in the bottom-right plot. The straight line indicates the charge-light correlation angle. A charge-light combined spectrum (bottom-left) shows a much improved energy resolution of 1.7\%($\sigma$).}
\label{ql_1kv}
\end{figure}

Fig. \ref{ql_1kv} shows the strong anti-correlation of charge and light signals measured with our detector for 662 keV $\gamma$-rays from $^{137}$Cs at 1 kV/cm. The energy resolution inferred from the light and the charge signals separately is 10.3\% ($\sigma$)  and 4.8\%($\sigma$), respectively. The resolution from the charge signal is consistent with previously measured values \cite{Aprile:01} and \cite{curioni:physics/0702078}. The charge-light correlation angle, $\theta$, also shown in Fig. \ref{ql_1kv} is defined as the angle between the major axis of the charge-light ellipse and the X-axis for light. $\theta$ can be roughly calculated as $\tan^{-1}(R_q/R_s)$, where $R_s$ and $R_q$ are the energy resolutions of the 662 keV peak from scintillation and ionization spectra separately. $\theta$ can also be found by a 2D gaussian fit to the charge-light ellipse of the 662 keV peak. A better energy resolution can be achieved by combining the charge and light signals as,
\begin{equation}
\label{eq:com}
\varepsilon_c = \frac{\sin{\theta}\cdot \varepsilon_s + \cos{\theta} \cdot \varepsilon_q}{\sin{\theta}+ \cos{\theta}}
\end{equation}
where $\varepsilon_c$ is the combined signal, in unit of keV. $\varepsilon_s$ and $\varepsilon_q$ are scintillation light and charge based energy in units of keV. The charge-light combined energy resolution of 662 keV line is significantly improved to 1.7\%($\sigma$).

The energy resolution from the charge-light combined spectrum, $R_c$, can be derived from equation \ref{eq:com} as \cite{Bevington},
\begin{equation}
\label{eq:comres}
R^2_c = \frac{\sin^2{\theta} \cdot R^2_s + \cos^2{\theta} \cdot R^2_q + 2\sin{\theta}\cos{\theta} \cdot R_{sq}}{(\sin{\theta}+ \cos{\theta})^2}
\end{equation}
where $R_s$ and $R_q$ are the energy resolutions from scintillation and ionization spectra separately. The covariance $R_{sq}$ is the contribution from the correlation of the two signals. The magnitude of $R_{sq}$ indicates the strength of anti-correlation (or correlation) between the scintillation and ionization signals. It is usually expressed in terms of correlation coefficient $\rho_{sq}$,
\begin{equation}
\label{eq:corcoe}
\rho_{sq} = R_{sq}/(R_sR_q)
\end{equation}
A value of $\rho_{sq}$ close to -1 (or 1) indicates a very strong anti-correlation (or correlation) of scintillation and ionization signals, while a zero $\rho_{sq}$ means no correlation. In equation \ref{eq:comres}, $R_s$ and $R_q$ can be expressed as,
\begin{equation}
\label{eq:lres}
{R_s}^2 = R_{si}^2 + R_{sg}^2 + R_{ss}^2 \approx R_{si}^2 + R_{ss}^2 
\end{equation}
\begin{equation}
\label{eq:qres}
{R_q}^2 \approx R_{qi}^2 + R_{qe}^2
\end{equation}
where $R_{si}$ and $R_{qi}$ are the energy resolution of scintillation and ionization, respectively contributed by liquid xenon itself. They include the liquid xenon intrinsic resolution and the contribution from fluctuations of electron-ion recombination. $R_{sg}$ is from the geometrical fluctuation of light collection. It is negligible in our result since only events in the center of the detector were selected for the analysis. $R_{ss}$ is from the statistical fluctuation of the number of photoelectrons $N_{pe}$ in the PMTs. $R_{ss}$ can be calculated roughly as $R_s = \sqrt{[1+(\frac{\sigma_g}{g})^2]/N_{pe}}$, which includes the statistical fluctuations of the number of photoelectrons and the PMTs gain variation ($\sigma_g/g \sim 0.67$, based on the single-photoelectron spectrum). $R_{qe}$ is from the noise equivalent charge ($ENC$) of the charge readout. $ENC$ was measured to be between 600 and 800 electrons, depending on the drift field, from a test pulse distribution. $R_{qe}=ENC/N_e$, where $N_e$ is the number of collected charges from the 662 keV peak. We note that we have neglected other contributions to the resolution of the charge measurement, such as from shielding grid inefficiency or pulse rise time variation, as they are sub-dominant compared to the electronic noise contribution. 

Table \ref{table1} lists the energy resolution of the 662 keV $\gamma$-ray peak inferred from ionization, scintillation, and charge-light combined spectra at different drift fields. The quoted errors are statistical only. The correlation angle and the correlation coefficient at each field are also presented. The energy resolution  improves with increasing field for both scintillation and ionization, while the charge-light combined energy resolution is about the same at different fields. The best value achieved in this study is 1.7\%($\sigma$) at 1 kV/cm drift field. We should mention that during this work, we observed improvement of the energy resolution from both light and combined energy spectra with improved light collection efficiency  by using Teflon, while the energy resolution from the charge spectrum did not change. 

\begin{table}[htbp]
\begin{center}
\begin{tabular}{cccccc}
\hline
Field [kV/cm] & $R_s(\%)$ & $R_q (\%)$ & $R_c(\%)$ & $\theta$ & $\rho_{sq}$ \\
\hline
0 & 7.9$\pm$0.3 &  & & \\
1 & 10.3$\pm$0.4 & 4.8$\pm$0.1 & 1.7$\pm$0.1 & 24.8$\rm^o$ & -0.87 \\
2 & 10.5$\pm$0.3 & 4.0$\pm$0.1 & 1.8$\pm$0.1 & 20.8$\rm^o$ & -0.80 \\
3 & 10.0$\pm$0.3 & 3.6$\pm$0.1 & 1.9$\pm$0.1 & 19.7$\rm^o$ & -0.74 \\
4 & 9.8$\pm$0.3 & 3.4$\pm$0.1 & 1.8$\pm$0.1 & 19.1$\rm^o$ & -0.74 \\
\hline
\end{tabular}
\caption{ {\it Measured} energy resolutions ($\sigma$) and correlation coefficients for 662 keV gamma rays at different electric field values.}
\label{table1}
\end{center}
\end{table}

The different value of the charge-light correlation coefficient at different fields indicates a more fundamental correlation coefficient between ionization and scintillation in liquid xenon. In fact, the energy resolution $R_c$ from charge-light combined signals comes from two factors. One is the liquid xenon intrinsic energy resolution $R_{ci}$. Another factor, $R_{ce}$, is contributed by external sources, such as the fluctuation of light collection efficiency on the light signal  and fluctuation of electronic noise on the charge signal. The charge-light combined energy resolution can be written as below,
\begin{equation}
\label{eq:comres2}
R^2_c  = R^2_{ci} + R^2_{ce} 
\end{equation}
\begin{equation}
\label{eq:comint}
R^2_{ci} = \frac{\sin^2{\theta} \cdot R^2_{si} + \cos^2{\theta} \cdot R^2_{qi} + 2\sin{\theta}\cos{\theta} \cdot R_{sqi}}{(\sin{\theta}+ \cos{\theta})^2} 
\end{equation}
\begin{equation}
\label{eq:comext}
R^2_{ce} \approx \frac{\sin^2{\theta} \cdot R^2_{ss} + \cos^2{\theta} \cdot R^2_{qe}}{(\sin{\theta}+ \cos{\theta})^2}
\end{equation}
In these equations $R_{si}$, $R_{qi}$ are the liquid xenon energy resolution from scintillation and ionization separately, as previously discussed. $R_{sqi}$ indicates the correlation between ionization and scintillation signals in liquid xenon. We can define the {\it intrinsic} correlation coefficient, $\rho_{sqi}$, of liquid xenon scintillation and ionization, similar to equation \ref{eq:corcoe} for the {\it measured} charge-light correlation coefficient, but without instrumental noise contributions.
\begin{equation}
\label{eq:corcoei}
\rho_{sqi} = R_{sqi}/(R_{si}R_{qi})
\end{equation}

The energy resolution for scintillation, $R_{si}$, and ionization, $R_{qi}$, can be calculated based on equation \ref{eq:lres} and \ref{eq:qres}, from the measured values of correlation angle $\theta$, statistical fluctuation of light detection $R_{ss}$ and electronic noise contribution $R_{qe}$. The calculated values are listed in Table \ref{table2}. Table \ref{table2} also shows the intrinsic and external contributions, $R_{ci}$ and $R_{ce}$, to the charge-light combined energy resolution. The values of $R_{ci}$ and $R_{ce}$ are calculated from equation \ref{eq:comres2}-\ref{eq:comext}. The intrinsic correlation coefficients from equation \ref{eq:corcoei} are also shown.
\begin{table}[htbp]
\begin{center}
\begin{tabular}{cccccc}
\hline
Field [kV/cm]  & $R_{si}$(\%) & $R_{qi}$(\%) & $R_{ce}$(\%) & $R_{ci}$(\%) & $\rho_{sqi}$ \\
\hline
0 & 6.0$\pm$0.3 &  & & \\
1 & 9.9$\pm$0.4 & 4.3$\pm$0.1 & 1.6$\pm$0.1 & $<1.0$ & $-1.00$ \\
2 & 10.1$\pm$0.3 & 3.5$\pm$0.1 & 1.7$\pm$0.2 & $<1.2$ & $-0.98$ \\
3 & 9.5$\pm$0.3 & 3.0$\pm$0.1 & 1.8$\pm$0.2 & $<1.2$  & $-0.98$ \\
4 & 9.3$\pm$0.3 & 2.8$\pm$0.1 & 1.8$\pm$0.2 & $<1.0$  & $-1.00$ \\
\hline
\end{tabular}
\caption{{\it Predicted} achievable energy resolutions in liquid xenon for light ($R_{si}$), charge ($R_{qi}$) and combined ($R_{ci}$) measurements, and charge-light correlation coefficient by removing instrumental noise contributions.}
\label{table2}
\end{center}
\end{table}

The intrinsic energy resolution in liquid xenon from the combined  scintillation and ionization signals is estimated to be less than 1\%.  Only  upper limits are given here since the uncertainties become large at such small values. The intrinsic correlation coefficients are closer to -1 than those measured from the experimental data including instrumental noise-contributions. This indicates near-perfect anti-correlation between ionization and scintillation in liquid xenon. We therefore expect that further improvement in the combined signal energy resolution can be achieved by increasing light collection efficiency and by minimizing  electronic noise. 

\subsection{Energy dependence of resolution}
\begin{figure}
\centering
\includegraphics[width=0.4\textwidth]{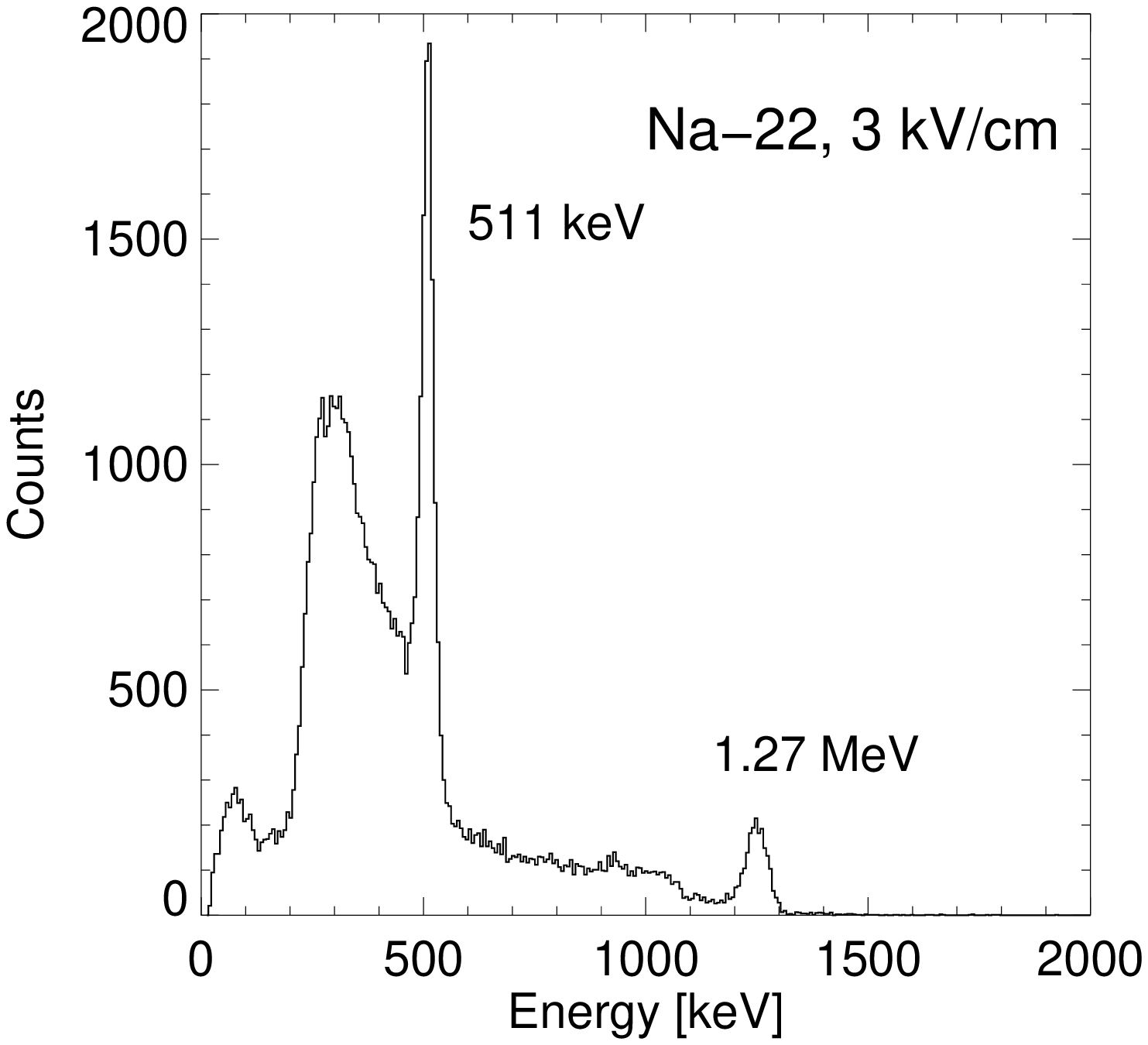}
\includegraphics[width=0.4\textwidth]{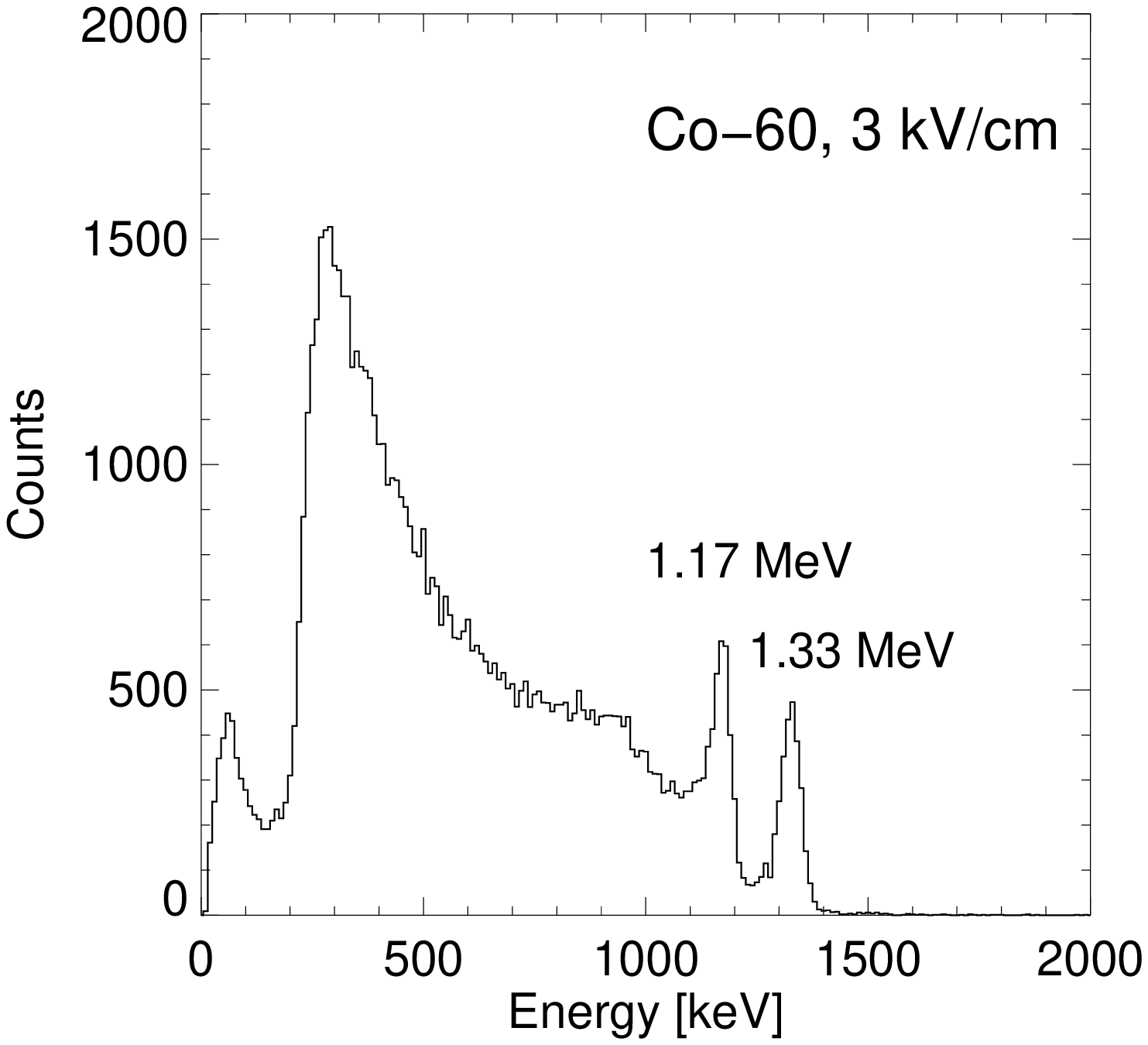}
\caption{Energy spectra of $^{22}$Na and $^{60}$Co $\gamma$ ray sources at 3 kV/cm, by combining charge and light signals.}
\label{Co60_Na22}
\end{figure}

\begin{figure}
\centering
\includegraphics[width=0.45\textwidth]{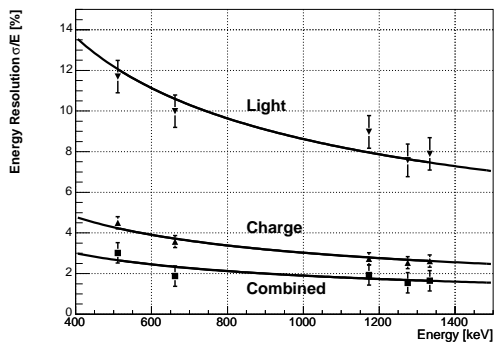}
\caption{Energy dependence of resolution measured from $^{22}$Na, $^{137}$Cs, and $^{60}$Co at 3 kV/cm drift field.}
\label{ene_3kv}
\end{figure}

The improvement of the energy resolution by combining scintillation and ionization signals was studied at 3 kV/cm drift field as a function of  $\gamma$-ray energy, using radioactive sources such as $^{22}$Na (511 keV and 1.28 MeV), $^{137}$Cs (662 keV) and $^{60}$Co (1.17 MeV and 1.33 MeV) (Fig. \ref{Co60_Na22}). The energy resolution from charge, light and charge-light combined spectra is shown in Fig.~\ref{ene_3kv}. The data was fitted with an empirical function, $\sigma/E = \alpha/\sqrt{(E/MeV)}$, yielding for the parameter $\alpha$ (8.6$\pm$0.4)\%, (3.0$\pm$0.4)\% and (1.9$\pm$0.4)\% for light, charge and combined spectra separately.

\section{Conclusion}

 We have shown that the energy resolution of MeV $\gamma$-rays in liquid xenon can be significantly improved by combining simultaneously measured scintillation and ionization signals. The best resolution achieved is  1.7\% ($\sigma$) for 662 keV $\gamma$-rays at 1 kV/cm. This value is much better than that measured from scintillation [10.3\% ($\sigma$)] or from ionization  [4.8\%($\sigma$)], separately. By summing the two signals, which are strongly anti-correlated, recombination fluctuations are reduced, resulting in improved energy resolution. At present, the energy resolution of the combined signal is still limited by external factors, such as light collection efficiency, PMT quantum efficiency and charge readout  electronic noise. By reducing the contribution from these factors, we estimate that the intrinsic energy resolution of MeV gamma-rays in liquid xenon from charge-light combined signal should  be less than 1\%. The simultaneous detection of ionization and scintillation signals in liquid xenon therefore provides a practical way to improve the energy measurement with a resolution better than the  Poisson limit, and possibly closer to the Fano limit \cite{Fano}. On the other hand, the limit to the energy resolution of LXe might well be not determined by Fano statistics but rather connected to the liquid phase, for instance to microscopic density non-uniformities of the liquid itself. It has been shown by Bolotnikov and Ramsey \cite{Bolotnikov:99} that the energy resolution deteriorates as the density of Xe gas increases. The behavior was attributed to the formation of molecular clusters in high pressure and liquid xenon, but a more quantitative explanation is needed.   Despite the limitation of energy resolution, the liquid phase offers far too many advantages for high energy radiation detection. The development of a LXeTPC which combines millimeter spatial resolution with  1\% or better energy resolution, within a large homogeneous volume, is very promising  for particle physics and astrophysics experiments. 
\section{Acknowledgments}
This work was supported by a grant
from the National Science Foundation (Grant No.  PHY-02-01740) to the Columbia Astrophysics Laboratory. The authors would like to express their thanks to Tadayoshi Doke and Akira Hitachi for valuable discussions.

\end{document}